\documentclass[aps]{revtex4}
\usepackage{graphicx}
\usepackage{amsmath}
\usepackage{amsfonts}
\usepackage{upgreek}
\usepackage{slashed}
\usepackage{color}
\usepackage{mathrsfs}
\usepackage{amssymb}
\newcommand{\vardelta}{\vartriangle\!\!}
\DeclareMathOperator{\thetaf}{\uptheta}
\DeclareMathOperator{\Tr}{Tr}

\DeclareMathOperator{\diag}{diag}
\begin{document}
\title{Multi Phase in Cold Dense Quark Matter}
\author{Song Shi$^1$}
\author{Juan Liu$^1$}
\author{Zhu-fang Cui$^2$}
\author{Hong-Shi Zong$^{2,3}$}\email[]{zonghs@nju.edu.cn}
\affiliation{$^{1}$ Department of Physics, National University of Defense Technology, Changsha 410000, China}
\affiliation{$^{2}$ Department of Physics, Nanjing University, Nanjing 210093, China}
\affiliation{$^{3}$ Joint Center for Particle, Nuclear Physics and Cosmology, Nanjing 210093, China}

\begin{abstract}
In this article, we study dynamic chiral symmetry breaking at zero temperature, finite chemical potential and external magnetic field with massless NJL model. We have proposed a mathematical method to classify phases in phase diagram of cold dense quark matter, and use mathematical analysis to identify the multi phase phenomenon among solutions for gap equation which means with fixed chemical potential and magnetic field, there could be two phases coexisting.

\bigskip


\end{abstract}

\maketitle

\section{Introduction}
The phase structure of QCD matter has always been an important and attractive topic in theoretical physics \cite{fukushima,costa,chao,zhao,jiang,xu,fu}. In relativistic heavy-ion collisions, the produced QCD matter will go though a phase transition or a crossover as time goes by. Either way, the state of QCD matter is believed to change from quark-gluon plasma to hadronic matter in this process. Its physical properties and dynamical behaviors such as chiral symmetry and confinement are altered along with the change of the state.

At the early stage of noncentral collision, the QCD matter produces extremely strong magnetic field \cite{kharzeev,kha,sko,vor}, which brings about obvious magnetic effects. Moreover, the compact stellar objects such as magnetars are believed had strong magnetic field around $10^{15}$G at their surface \cite{dun,kou}, Therefore studying QCD matter's properties under the influence of magnetic field becomes a meaningful and important subject. So far, many relevant theories and models have been proposed and it is shown that the quark condensate are strengthened by magnetic field, which is known as `Magnetic Catalysis' \cite{gusynin,gusynin2,ebert,shov,alan}. Consequently, the QCD phase diagram is related to magnetic field \cite{andersen,endrodi}.

In this article, we will study the phase property of so called ``cold dense quark matter'', unlike the quark matter in high energy experiments, this matter has low temperature, hence we can establish models at zero temperature limitation. This kind of research could facilitate the study of compact stellar objects. Although similar projects have been thoroughly studied in early articles \cite{ebert2,ebert3,ina,men,boom,faya,ferr,ferrer,mand,mand2,pre,pre2,allen}, but in this article we try to study phase diagram of cold dense matter in a different point of view, we have developed a mathematical method to classify phases in phase diagram, and find out new property from gap equation.

The model we employ is the two-flavor NJL model at chiral limitation with mean-field approximation \cite{vogl,klev,hats,bub}, it is a good tool to investigate dynamical chiral symmetry breaking of nonperturbative QCD matter. One thing need mentioning here, in Asakawa's work \cite{asa}, it was pointed that with the presence of chemical potential, the self-energy does not simply equal dynamical mass, which reveals with the help of the Fierz transformation. The actual self-energy should be written as $\Sigma=\sigma+a\gamma^0$ to guarantee the self-consistency of gap equation. In our case, this problem becomes much more complicate, the external magnetic field and chemical potential render self energy has four kinds of mean fields, $\Sigma=\sigma+a\gamma^0+b\gamma^5\gamma^3+c\sigma^{12}$. At zero temperature limitation, this will cause very chaotic situation in gap equation. But fortunately, except $\sigma$, the other three quantities are very small comparing to nonzero solutions of $\sigma$, we can ignore them for a schematic view of phase diagram.

The article is arranged as below, Section \ref{ge} is the deduction of gap equation with our new developed method different from Schwinger's ``proper time'', this method could handle more complicate models or ansatz such as $\Sigma=\sigma+a\gamma^0+b\gamma^5\gamma^3+c\sigma^{12}$, one can refer to Appendix \ref{prop} and \ref{eigen} for more details of this method. Section \ref{analysis} is the classification of phase diagram by using mathematical analysis and numerical analysis, in this section we mathematically define several areas in phase diagram which have different phase properties. Section \ref{concl} is the conclusion.

\section{Gap Equation\label{ge}}
We start from the lagrangian below (which is the NJL model with mean field approximation)
\begin{equation}
 \mathcal{L}=\bar\psi(\slashed{D}+\mu\gamma^0-\sigma)\psi-\frac{N_\text{c}}{2G}\sigma^2,\label{la0}
\end{equation}
\begin{equation}
 D_\mu=i\partial_\mu+eA_\mu\otimes Q,\qquad  A_\mu=(0,\frac{B}{2}x^2,-\frac{B}{2}x^1,0),\qquad Q=\diag(q_\text{u}, q_\text{d}),
\end{equation}
here $A_\mu$ gives us the external magnetic field with the strength $B$ that parallels to $x^3$ axis. $q_\text{u}$, $q_\text{d}$ separately represent electric charge numbers of up quark and down quark, hence $q_\text{u}=\frac{2}{3}$, $q_\text{d}=-\frac{1}{3}$. We employ $q_\text{f}$ to generally represent $q_\text{u}$, $q_\text{d}$ in following discussions, and the index `f' can be assigned to `u' or `d'.

In the Lagrangian Eq. (\ref{la0}) we have employed the ansatz that $\Sigma=\sigma I_4$, while actually, with nonzero external magnetic field and chemical potential, $\Sigma$ should have not only quark condensate but also vector, axial vector and tensor condensates. The last there kinds of condensates are thought to be too small to affect the general properties of QCD matter but some subtleties, therefore in this paper we assume $\Sigma=\sigma I_4$.

Now we manage to get the fermion propagator with magnetic field and chemical potential, in the mean time, to identify the appropriate form of $\varepsilon$ term (because it is zero temperature). According to Appendix A, in order to acquire the correct fermion propagator simply, one has to multiply Hamiltonian density with a factor $(1-i\eta)$,
\begin{equation}
 \mathcal{H}'=-(1-i\eta)\bar\psi(\gamma^iD_i+\mu\gamma^0-\sigma)\psi+\frac{N_\text{c}}{2G}\sigma^2(1-i\eta).
\end{equation}
Consequently, the Lagrangian has changed to,
\begin{equation}
 \mathcal{L}'=\sum_\text{f}\bar\psi\hat S_\text{f}'^{-1}\psi-\frac{N_\text{c}}{2G}\sigma^2(1-i\eta),\qquad\hat S'_\text{f}=\frac{1}{/\kern-0.55em\hat\Pi^\text{f}-\sigma'},\label{la1}
\end{equation}
\begin{equation}
 \sigma'=(1-i\eta)\sigma,\quad\hat\Pi^\text{f}_\mu=\hat p'_\mu+q_\text{f}eA_\mu(1-i\eta),\quad\hat p'_0=\hat p_0+(1-i\eta)\mu,\quad\hat p'_i=(1-i\eta)\hat p_i,
\end{equation}
the $\sum_\text{f}$ here means we have separated flavor space, and the fermion field operator $\psi$ in Eq. (\ref{la1}) is $4$ components single flavor spinor rather than $8$ components two flavor spinor in Eq. (\ref{la0}).

Through the definition of partition function
\begin{equation}
 \mathcal{Z}=\lim_{\eta\to0^+}\int\mathrm{D}\bar\psi\,\mathrm{D}\psi\,e^{i\int\mathcal{L}'\,\mathrm{d}^4x}=\lim_{\eta\to0^+}e^{-iW'(\sigma,\mu,eB,\eta)},
\end{equation}
we have the effective action $W'$
\begin{equation}
 W'=\frac{N_\text{c}}{2G}\sigma^2(1-i\eta)\int\mathrm{d}^4x+iN_\text{c}\sum_\text{f}\Tr(\ln\hat S_\text{f}'^{-1}).\label{free}
\end{equation}
The principle of gap equation is to identify the least or local minimum value for effective action, which is
\begin{equation}
 \frac{\delta W'}{\delta\sigma}=0,
\end{equation}
hence we have
\begin{equation}
 (1-i\eta)\frac{\sigma}{G}\int\mathrm{d}^4x=i\sum_\text{f}\Tr\hat S'_\text{f}.\label{gap1}
\end{equation}
in following discussion we can safely set factor $(1-i\eta)$ to be $1$ at the left hand side of Eq. (\ref{gap1}) with the limitation $\eta\to0^+$.

Now we need to take care of $\Tr\hat S'_\text{f}$ in Eq. (\ref{gap1}), because there is chemical potential in the propagator, it is not convenient to use Schwinger's `proper time' method. We have developed a new method to deal such situation. One can refer to our previous work \cite{shi} for a detailed introduction or to Appendix B for an overview.

According to Eq. (\ref{gapb}), we have
\begin{equation}
 \Tr\hat S'_\text{f}=\frac{|q_\text{f}|eB\sigma}{\pi}\int\frac{\mathrm{d}p_0\,\mathrm{d}p_3}{(2\pi)^2}\,
 \sum_n\frac{2-\delta_{0n}}{p'^2_0-2n|q_\text{f}|eB'-p'^2_3-\sigma'^2}\int\mathrm{d}^4x,\label{trace1}
\end{equation}
\begin{equation}
 p'_0=p_0+(1-i\eta)\mu,\qquad p'_3=(1-i\eta)p_3,\qquad B'=(1-i\eta)^2B.
\end{equation}
Noticing, the complex factor in $B'$ is $(1-i\eta)^2$ rather than $(1-i\eta)$, because $2n|q_\text{f}|eB'$ comes from the combination of quantization of $\hat\Pi^2_\perp$ and $q_\text{f}eB\sigma^{12}$ which is the outcome of $[\hat\Pi^\text{f}_1,\hat\Pi^\text{f}_2]$, both of them provide the factor $(1-i\eta)^2$. Secondly one should be aware due to nonzero chemical potential, $\Tr(\hat S'_\text{f}\gamma^0)$ is not $0$, the consequence is that we should introduce a shift for chemical potential to prevent the inconsistency. But as we have mentioned at the beginning, such shift relates to a vector condensate and it is very small, we could exclude its effect in our qualitative results.

Continuing to adjust the expression of Eq. (\ref{trace1})
\begin{equation}
 \begin{split}
  \Tr\hat S'_\text{f}&=\frac{|q_\text{f}|eB\sigma}{\pi}\int\frac{\mathrm{d}p_0\,\mathrm{d}p_3}{(2\pi)^2}\,
  \sum_n\frac{2-\delta_{0n}}{(p_0+\mu)^2-\omega_{n\text{f}}^2+i\varepsilon[\omega_{n\text{f}}^2-\mu(p_0+\mu)]}\int\mathrm{d}^4x\\
  &=\frac{|q_\text{f}|eB\sigma}{\pi}\int\frac{\mathrm{d}p_0\,\mathrm{d}p_3}{(2\pi)^2}\,
  \sum_n\frac{2-\delta_{0n}}{p_0-\omega_{n\text{f}}^2+i\varepsilon(\omega_{n\text{f}}^2-\mu p_0)}\int\mathrm{d}^4x,
 \end{split}\label{trace2}
\end{equation}
\begin{equation}
 \omega_{n\text{f}}=\sqrt{p_3^2+2n|q_\text{f}|eB+\sigma^2},\qquad \varepsilon=2\eta\to0^+.
\end{equation}
Making a Wick rotation to Eq. (\ref{trace2}), and applying proper time method
\begin{equation}
 \frac{\Tr\hat S'_\text{f}}{\int\mathrm{d}^4x}=-i\frac{|q_\text{f}|eB\sigma}{4\pi^2}\int_0^{+\infty}\frac{e^{-\sigma^2s}}{s}\coth(|q_\text{f}|eBs)\,\mathrm{d}s
 +i2\sum_\text{f}|q_\text{f}|eB\sigma\sum_n(2-\delta_{0n})\thetaf(\mu-\lambda_{n\text{f}})\ln\frac{\mu+\sqrt{\mu^2-\lambda_{n\text{f}}^2}}{\lambda_{n\text{f}}},\label{trace3}
\end{equation}
\begin{equation}
 \lambda_{n\text{f}}=\sqrt{\sigma^2+2n|q_\text{f}|eB}.
\end{equation}

Combining Eqs. (\ref{gap1}) and (\ref{trace3}), making a truncation to proper time `$s$' ($\int_0^{+\infty}\mathrm{d}s\to\int_{1/\Lambda^2}^{+\infty}\mathrm{d}s$), the gap equation could be simplified to (despite the trivial solution $\sigma=0$)
\begin{equation}
 \begin{split}
  \frac{4\pi^2}{G}=&\sum_\text{f}|q_\text{f}|eB\int_{1/\Lambda^2}^{+\infty}\frac{e^{-\sigma^2s}}{s}\coth(|q_\text{f}|eBs)\,\mathrm{d}s\\
  &-2eB\thetaf(\mu-\sigma)\ln\frac{\mu+\sqrt{\mu^2-\sigma^2}}{\sigma}-4\sum_\text{f}|q_\text{f}|eB\sum_{n=1}^{+\infty}\thetaf(\mu-\lambda_{n\text{f}})
  \ln\frac{\mu+\sqrt{\mu^2-\lambda_{n\text{f}}^2}}{\lambda_{n\text{f}}},
 \end{split}\label{gap2}
\end{equation}

At the $B\to0^+$ and $\mu\to0^+$ limits, Eq. (\ref{gap2}) degenerates to the classic gap equation
\begin{equation}
 \frac{4\pi^2}{G}=2\int_{1/\Lambda^2}^{+\infty}\frac{e^{-\sigma^2s}}{s^2}\mathrm{d}s,
\end{equation}
which we can use to determine the value of $\Lambda$ and $G$ \cite{inagaki},
\begin{equation}
 \Lambda=0.99\text{GeV},\qquad G=25.4\text{GeV}^{-2}.
\end{equation}

\section{Analysis and Numerical Results\label{analysis}}

\subsection{The Boundary of Chemical Potential}
There is a boundary to chemical potential. Before explain that, we have to vary gap equation Eq. (\ref{gap2}) a little firstly
\begin{eqnarray}
 f(\sigma,eB)&=&h(\sigma,\mu,eB),\nonumber\\
 f(\sigma,eB)&=&\sum_\text{f}|q_\text{f}|eB\int_{1/\Lambda^2}^{+\infty}\frac{e^{-\sigma^2s}}{s}\coth(|q_\text{f}|eBs)\,\mathrm{d}s,\nonumber\\
 h(\sigma,\mu,eB)&=&2eB\thetaf(\mu-\sigma)\ln\frac{\mu+\sqrt{\mu^2-\sigma^2}}{\sigma}+4\sum_\text{f}|q_\text{f}|eB\sum_{n=1}^{+\infty}\thetaf(\mu-\lambda_{n\text{f}})
  \ln\frac{\mu+\sqrt{\mu^2-\lambda_{n\text{f}}^2}}{\lambda_{n\text{f}}}+\frac{4\pi^2}{G}\label{gap3}
\end{eqnarray}
The first row is the variation. In this way, we can study two functions $f(\sigma,eB)$ and $h(\sigma,\mu,eB)$ separately beside gap equation.

\begin{figure}
 \centering
 \includegraphics[width=3in]{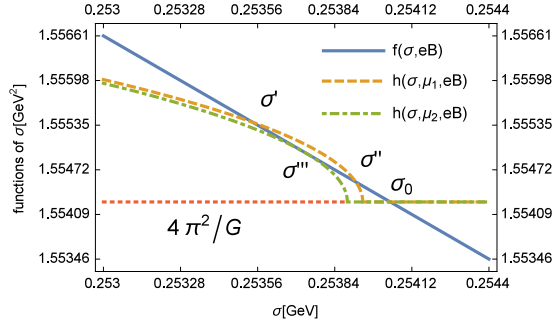}
 \caption{In this figure, we treat $\sigma$ as variable, $eB$ and $\mu$ as preset parameters, and the vertical axis as function of $\sigma$. $eB=0.01$GeV$^2$, $\mu_1\approx0.25394$GeV, $\mu_2\approx0.253886$GeV. The horizontal dotted line represents $\frac{4\pi^2}{G}$. $f(\sigma,eB)$ intersects with $h(\sigma,\mu_1,eB)$ at there points, $\sigma'$, $\sigma''$ and $\sigma_0$. $f(\sigma,eB)$ contacts with $h(\sigma,\mu_2,eB)$ at $\sigma'''$.\label{elu1}}
\end{figure}

From the expression of $h(\sigma,\mu,eB)$ in Eq. (\ref{gap3}) we know if chemical potential $\mu$ is smaller than dynamic mass $\sigma$, then the gap equation is simplified to
\begin{equation}
 f(\sigma,eB)=\frac{4\pi^2}{G},\label{gap0}
\end{equation}
there will be no $\mu$-dependent dynamic mass in existence, we define these ordinary `only-magnet-dependent' dynamic mass (ODM) as $\sigma_0$. $\sigma_0$ roughly gives us lower boundary of $\mu$ that the solutions of gap equation Eq. (\ref{gap3}) are beyond `ordinary'. In the matter of fact, the actual lower boundary is generally smaller than $\sigma_0$ with all $eB$s, take Fig. (\ref{elu1}) for example, we have three solutions for gap equation in the case of $\mu_1$, $\sigma_0$ is the ODM, $\sigma'$ is also a valid solution for Eq. (\ref{gap3}) while $\sigma''$ is not, because from mathematical analysis we know at $\sigma''$ the effective action Eq. (\ref{free}) has a local maximum rather than minimum, it is not what we need. In following discussion we will ignore these maximum points. The contacting points of $f(\sigma,eB)$ and $h(\sigma,\mu,eB)$ in Fig. (\ref{elu1}) is not a valid solution neither, but it is a boundary point for chemical potential. Comparing $\mu_1$ line with $\mu_2$, we can see when $\mu>\mu_2$, Eq. (\ref{gap3}) has valid solutions beyond ODM, and analytically it is true, because $h(\sigma,\mu,eB)$ is a monotonically increasing function by $\mu$, as long as $\mu$ is bigger than a specific value at different $eB$s, the solution will be not just ODM. Therefore the real lower boundary of $\mu$, defined as $\mu_\text{low}$, is always lower than $\sigma_0$.

To identify $\mu_\text{low}$, we have to solve the simultaneous equations of $\sigma$,
\begin{equation}
 f(\sigma,eB)|_{\sigma=\mu_\text{low}}=h(\sigma,\mu,eB)|_{\sigma=\mu_\text{low}},
 \qquad\left.\frac{\partial f(\sigma,eB)}{\partial\sigma}\right|_{\sigma=\mu_\text{low}}=\left.\frac{\partial h(\sigma,\mu,eB)}{\partial\sigma}\right|_{\sigma=\mu_\text{low}}.\label{mulow}
\end{equation}

Beside lower boundary, we also have upper boundary for chemical potential, defined as $\mu_\text{up}$, when $\mu$ exceeds such boundary, Eq. (\ref{gap3}) has no valid solution ($f(\mu,eB)$ has no intersection with $h(\sigma,\mu,eB)$, or the intersection point represents maximum rather minimum of effective action), which means chiral symmetry restores, Wigner phase arises. How to identify the upper boundary depends on the asymptotic behaviors of $f(\mu,eB)$ and $h(\sigma,\mu,eB)$ at $\sigma\to0^+$. First of all, when $\sigma$ is big enough, we will eventually have $h(\sigma,\mu,eB)>f(\sigma,eB)$, because $h(+\infty,\mu,eB)=4\pi^2/G$ and $f(+\infty,eB)=0$, so if $h(0^+,\mu,eB)<f(0^+,eB)$, by Bolzano's Theorem, there must be a valid solution of $\sigma$ in somewhere between $0$ and $+\infty$. But what if $h(0^+,\mu,eB)>f(0^+,eB)$? At $\sigma\to0^+$ and $\sigma\to+\infty$, we both have $h(\sigma,\mu,eB)>f(\sigma,eB)$, it seems using mathematica analysis to identify the intersections is impossible. Nevertheless, we should find out the asymptotic properties at $\sigma\to0^+$ first.

For the function $f(\sigma,eB)$,
\begin{equation}
 f(\sigma,eB)=\sum_\text{f}|q_\text{f}|eB\int_{1/\Lambda^2}^{+\infty}\frac{e^{-\sigma^2s}}{s}[\coth(|q_\text{f}|eBs)-1]\,\mathrm{d}s
 +eB\int_{1/\Lambda^2}^{+\infty}\frac{e^{-\sigma^2s}}{s}\,\mathrm{d}s,
\end{equation}
by L'H\^opital's rule,
\begin{equation}
 \lim_{\sigma\to0^+}\frac{\int_{1/\Lambda^2}^{+\infty}e^{-\sigma^2s}/s\,\mathrm{d}s}{\ln\sigma}
 =\lim_{\sigma\to0^+}\frac{\int_{\sigma^2/\Lambda^2}^{+\infty}e^{-s}/s\,\mathrm{d}s}{\ln\sigma}=-2,
\end{equation}
therefore we have
\begin{equation}
 \begin{split}
  f(\sigma,eB)&\sim f_0(\sigma,eB),\qquad\sigma\to0^+,\\
  f_0(\sigma,eB)&=-2eB\ln\sigma+eBC
  +\sum_\text{f}|q_\text{f}|eB\int_{1/\Lambda^2}^{+\infty}\frac{\coth(|q_\text{f}|eBs)-1}{s}\,\mathrm{d}s,\\
  C&=\lim_{\sigma\to0^+}\left(\int_{1/\Lambda^2}^{+\infty}\frac{e^{-\sigma^2s}}{s}\,\mathrm{d}s+2\ln\sigma\right)\approx-0.595297,\label{asym1}
 \end{split}
\end{equation}
here `$\sim$' reads ``$f(\sigma,eB)$ is asymptotic to $f_0(\sigma,eB)$ as $\sigma$ tends to $0^+$" \cite{bender}.

If someone get confused on the units problem in the logarithm function of $\sigma$, one can use
\begin{equation}
 -2\ln\sigma+C=-2\ln\frac{\sigma}{C'},\qquad C'\approx0.742562\text{GeV},
\end{equation}
to replace the one in Eq. (\ref{asym1}).

For the function $h(\sigma,\mu,eB)$,
\begin{equation}
 \begin{split}
  h(\sigma,\mu,eB)&\sim h_0(\sigma,\mu,eB),\qquad\sigma\to0^+,\\
  h_0(\sigma,\mu,eB)&=-2eB\ln\frac{\sigma}{2\mu}
  +4\sum_\text{f}|q_\text{f}|eB\sum_{n=1}^{+\infty}\thetaf(\mu-\lambda^0_{n\text{f}})\ln\frac{\mu+\sqrt{\mu^2-(\lambda^0_{n\text{f}})^2}}{\lambda^0_{n\text{f}}}
  +\frac{4\pi^2}{G},\nonumber\\
  \lambda^0_{n\text{f}}&=\sqrt{2n|q_\text{f}|eB}.
 \end{split}
\end{equation}
Noticing, as $\sigma\to0^+$, both $f(\sigma,eB)$ and $h(\sigma,\mu,eB)$ have the same dominant asymptotic behaviors, closing to $(-2eB\ln\sigma)$, this renders quite complicate relations between them, and the solutions too, when they are close enough. But still, we can make a use of their asymptotic behaviors to `roughly' identify the upper boundary of chemical potential by equation of $\mu$,
\begin{equation}
 f_0(\sigma,eB)=h_0(\sigma,\mu,eB)|_{\mu=\mu_\text{up}},\label{muup}
\end{equation}
the $\ln\sigma$ terms in Eq. (\ref{muup}) are perfectly canceled on both side, which leaves us an implicit function with respect to $\mu_\text{up}$ and $eB$.

\begin{figure}
 \centering
 \includegraphics[width=3in]{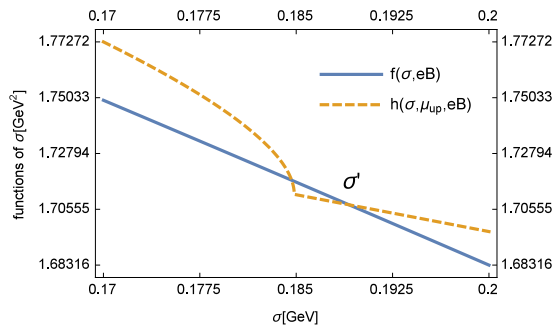}
 \caption{$eB=0.05$GeV$^2$, $\mu_\text{up}\approx0.317547$GeV. $\sigma'$ is a valid intersection.\label{inter1}}
\end{figure}

In the above discussion we have mentioned ``roughly identify the upper boundary", that's because the actual upper boundary is beyond Eq. (\ref{muup}). There are two kinds of situations. First, $f(\sigma,eB)$ is a smooth function of $\sigma$ but $h(\sigma,\mu,eB)$ is not, beside that, $f(\sigma,eB)$ and $h(\sigma,\mu_\text{up},eB)$ are not only very close as $\sigma\to0^+$, but also close enough when $\sigma$ stretches to a finite value, say $0.1$GeV, these two reasons cause multiple intersections before they distinctly separate, and of cause some of the intersections account for nonzero dynamic mass, e.g. Fig. (\ref{inter1}). But that doesn't say Eq. (\ref{muup}) is of no use, we can add a tiny modification to $\mu_\text{up}$ from Eq. (\ref{muup}), which brings us the actual upper boundary $\mu_\text{ub}=\mu_\text{up}+\!\vardelta\mu$, in the case of Fig. (\ref{inter1}), $\vardelta\mu\approx0.0018$GeV, it is a small quantity, therefore we can say Eq. (\ref{muup}) roughly identifies upper boundary.

\begin{figure}
 \centering
 \includegraphics[width=3in]{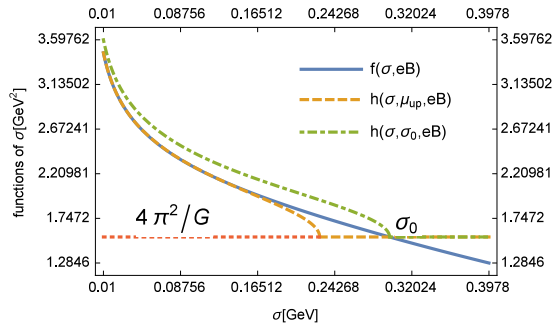}
 \caption{$eB=0.25$GeV$^2$, $\mu_\text{up}\approx0.227481$GeV, $\sigma_0\approx0.2978$GeV. $\sigma_0$ is a valid intersection between $f(\sigma,eB)$ and $h(\sigma,\mu_\text{up},eB)$, but is not valid between $f(\sigma,eB)$ and $h(\sigma,\sigma_0,eB)$.\label{inter2}}
\end{figure}

The second situation, through numerical result, we find that $\mu_\text{up}$ decreases along with increasing magnetic field, while the ODM $\sigma_0$ is increasing, therefore when $\mu_\text{up}$, as the function of $eB$, exceeds a threshold, $f(\sigma,eB)$ will always intersect with $h(\sigma,\mu_\text{up},eB)$ at a ODM point, e.g. Fig. (\ref{inter2}), this implies that the actual upper boundary, defined as $\mu_\text{ub}$, depends on $\sigma_0$ rather than $\mu_\text{up}$.

There is an equation to summarize above discussions,
\begin{equation}
 \mu_\text{ub}=\max(\sigma_0,\mu_\text{up}+\!\vardelta\mu).
\end{equation}

\begin{figure}
 \centering
 \includegraphics[width=3in]{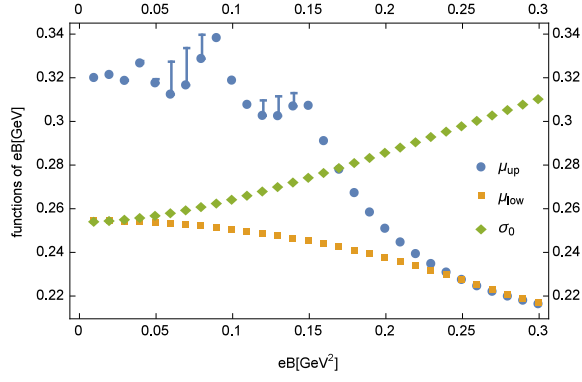}
 \caption{$\sigma_0$, $\mu_\text{up}$, $\mu_\text{low}$ are implicit functions of $eB$. The error bars of $\mu_\text{up}$ represent modification $\vardelta\mu$ to $\mu_\text{up}$, they are only sketch, the actual modifications are relatively small quantities. When $eB>0.16$GeV$^2$, $\mu_\text{up}<\sigma_0$, the actual upper boundary of chemical potential depends on $\sigma_0$ rather than $\mu_\text{up}$. Along with $eB$'s increasing, $\mu_\text{low}$ and $\mu_\text{up}$ are decreasing, and getting closer and closer, this property could be easily understood through mathematics analysis. Noticing, when $eB$ ranges from $0$ to $0.15$GeV$^2$, $\mu_\text{up}$ has irregular oscillation.\label{bound}}
\end{figure}

\begin{figure}
 \centering
 \includegraphics[width=3in]{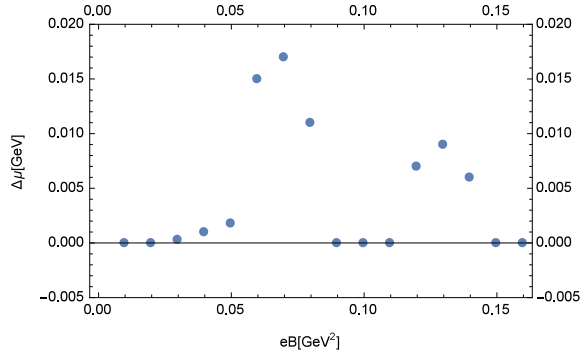}
 \caption{The modifications of $\mu_\text{up}$, lead to actual upper boundaries of chemical potential. The range of $eB$ is from $0$ to $0.16$GeV$^2$, because when $eB>0.16$GeV$^2$, the actual upper boundaries is determined by $\sigma_0$, there is no need to consider modification for $\mu_\text{up}$. The biggest modification happens at $eB=0.07$GeV$^2$, $\vardelta\mu\approx0.017$GeV, while the smallest modification, except $0$, is $\vardelta\mu\approx0.0003$GeV at $eB=0.03$GeV$^2$. At some points, $\vardelta\mu=0$, no modification is needed, $\mu_\text{ub}=\mu_\text{up}$.\label{error}}
\end{figure}

So far, we have discussed the properties of $\sigma_0$, $\mu_\text{low}$ and $\mu_\text{up}$, we put them in Fig. (\ref{bound}). The modifications of $\mu_\text{up}$ is also sketchily plotted in this figure as error bars, a precise demonstration of these modifications is shown in Fig. (\ref{error}), which, as we can see, is relatively small and ruleless, therefore for a qualitative discussion, we can just talk about $\mu_\text{up}$ instead of $\mu_\text{ub}$ when $\mu_\text{ub}=\mu_\text{up}+\!\vardelta\mu$. Noticeably, when magnetic field is not strong enough, relatively, the upper boundary of chemical potential, which separates chiral restored phase and chiral broken phase, is highly ruleless, it has irregular oscillation, but when $eB$ keeps increasing, $\mu_\text{up}$ begins regularly decreasing, we will discuss this in the conclusion.

\begin{figure}
 \centering
 \includegraphics[width=3in]{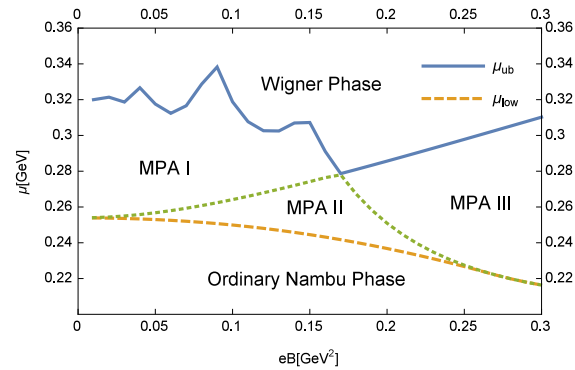}
 \caption{The phase diagram of $eB$ and $\mu$. Below the dashed line is ordinary Nambu phase, the dynamic mass is ODM, determined by Eq. (\ref{gap0}). Above the solid line is Wigner phase, $\sigma=0$. In between is the multi phase area (MPA), and it is separated to three areas by dotted line, the function for dotted line is $\min(\mu_\text{up},\sigma_0)$. These areas have different properties of dynamic mass.\label{phase}}
\end{figure}

From Fig. (\ref{bound}), the three functions of $eB$ have roughly divided the diagram into several areas, this gives us phase diagram with nonzero chemical potential and magnetic field at zero temperature, as shown in Fig. (\ref{phase}). The diagram is mainly divided to three kinds of areas, the ordinary Nambu phase, the multi phase area and the Wigner phase, and in multi phase area (MPA), it is divided into three subareas by function $\min(\mu_\text{up},\sigma_0)$. The ordinary Nambu phase has only one solution, the ODM, determined by Eq. (\ref{gap0}), it depends only on magnetic field, has no relationship with chemical potential. The Wigner phase is the chiral restored phase, it always has zero dynamic mass $\sigma=0$ in this model (chiral limit NJL model) we study. The most interesting part is the MPA, its three subareas have different properties of the solutions of gap equation Eq. (\ref{gap3}). We will discuss this in detail next subsection.

\subsection{The Solutions in Multi Phase Area}
In order to clarify the phase properties in MPA of Fig. (\ref{phase}), we have to treat $h(\sigma,\mu,eB)$ in a proper way. Due to the Heaviside step function, $h(\sigma,\mu,eB)$, as a function of $\sigma$, is continue but not smooth, it has stages, and at the lowest stage, $h$ becomes a constant $4\pi^2/G$, therefore we consider $h(\sigma,\mu,eB)$ as a two sections function, section I, the stairs ($h>4\pi^2/G$), in this section, $h$ could have many stairs, inside the stair, $h$ is smooth, at the point of two stairs contacting, $h$ is continue but not smooth, the $h$ function in Fig. (\ref{inter1}) is a good example, section II, the ground ($h=4\pi^2/G$), of cause at the point of stairs and ground contacting, $h$ is continue but not smooth, too. When $f$ and $h$ intersect at the ground, we have ODM, when at the stairs, we have UDM (unordinary dynamic mass).

\begin{figure}
 \centering
 \includegraphics[width=3in]{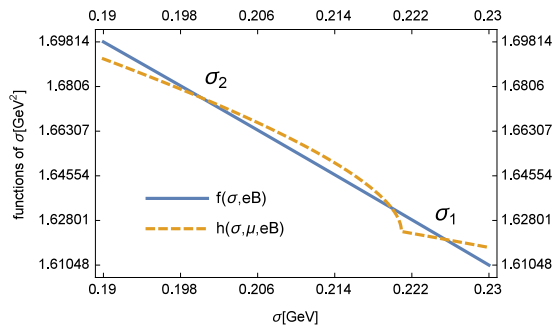}
 \caption{$eB=0.03$GeV$^2$, $\mu=0.298$GeV. Both $\sigma_1$ and $\sigma_2$ are valid solutions for gap equation.\label{inter3}}
\end{figure}

\begin{figure}
 \centering
 \includegraphics[width=3in]{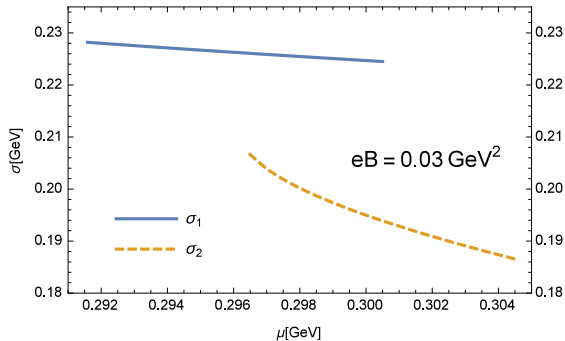}
 \caption{Defining $\vardelta\sigma=\sigma_1-\sigma_2$. The maximum of $\vardelta\sigma$ is approximately $0.03$GeV at $\mu\approx0.3005$GeV. The minimum of $\vardelta\sigma$ is approximately $0.019$GeV at $\mu\approx0.2965$GeV. $\mu\in(0.2965, 0.3005)$ is one multi phase interval for $\mu$ at $eB=0.03$GeV$^2$.\label{sigmu1}}
\end{figure}

MPA I is a area that only has UDMs. It quite clear that in this area no ODM solution is allowed, because $\mu>\sigma_0$. In MPA I, we also have $\mu>\mu_\text{low}$, $\mu_\text{low}$ means the point of first contact of $f$ and $h$, when $\mu$ exceeds the first contact point, $f$ and $h$ will have at least one intersection, of cause an UDM, or multi UDMs simultaneously, e.g. Fig. (\ref{inter3}). But multi UDMs does not happens all the time, generally, when the solution of gap equation is closing to a contacting point of two stairs of $h$, we can find another valid solution in the other stair. The $\sigma$-$\mu$ relation could be like Fig. (\ref{sigmu1}), which is just a small part of a big picture. When $\mu$ ranges from $\mu_\text{low}$ to $\mu_\text{up}$, the cascade could happen many times, and the multi phase areas only exist for a short range.

\begin{figure}
 \centering
 \includegraphics[width=3in]{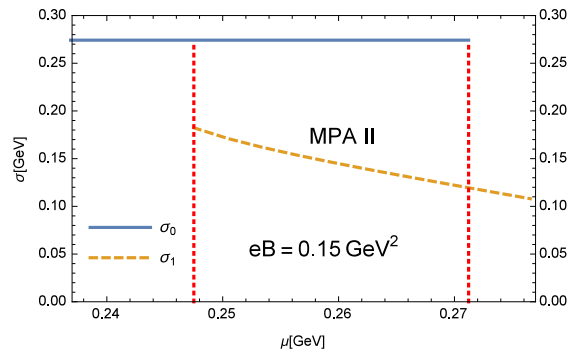}
 \caption{$\sigma_0$ is the ODM at $eB=0.15$GeV$^2$, $\sigma_1$ is the UDM. The dotted lines divide MPA II from other areas.\label{sigmu2}}
\end{figure}

MPA II is the area that definitely has multi phase simultaneously, in this area, we always have at least two valid solutions from the gap equation, and one of which is an ODM. Because $\mu<\sigma_0$, $f$ has an intersection with $h$ at the ground, while $\mu_\text{low}<\mu<\mu_\text{up}$, $f$ always cuts through the stairs of $h$, that brings us another valid solution. The $f$ and $h$'s relation is quite like Fig. (\ref{inter3}), except the lowest stair of $h$ is ground $4\pi^2/G$. The $\sigma$-$\mu$ relation about this area is shown in Fig. (\ref{sigmu2}).

MPA III is just another ordinary Nambu phase area, in this area, all dynamic mass are determined by Eq. (\ref{gap0}). From Fig. (\ref{phase}), it seems MPA III and ordinary Nambu phase from the bottom are separated by MPA II, but considering $\mu_\text{up}$ and $\mu_\text{low}$ are moving really closer with increasing $eB$, maybe they will connect when $eB$ is strong enough.

\section{The Conclusion\label{concl}}
In this article, we have developed a mathematical analysis method to help drawing phase diagram of cold dense quark matter, and roughly divide phase diagram into several areas, each area has a unique phase property. The phase diagram and division depend on three kinds of quantities, $\sigma_0$, $\mu_\text{low}$, $\mu_\text{up}$, they are all implicit functions of $eB$, the equations that depict these quantities are separately Eqs. (\ref{gap0}), (\ref{mulow}), (\ref{muup}). $\sigma_0$ is the ordinary dynamic mass in NJL model with external magnetic field, when chemical potential is involved, there are some dynamic mass deviating ODM, therefore $\sigma_0$ is the base line of all other phases. $\mu_\text{low}$ is the first contacting point of $f(\sigma,eB)$ and $h(\sigma,\mu,eB)$, when chemical potential is smaller than $\mu_\text{low}$, $f$ and $h$ can only have the ODM solutions for gap equation, therefore $\mu_\text{low}$ is the dividing line of ordinary dynamic mass and unordinary dynamic mass. The meaning of $\mu_\text{up}$ is a bit complicate, generally speaking, it is the dividing line of multi phase area and simple phase area, when magnetic field is relatively weak, say $0.15$GeV$^2$ comparing to $0.3$GeV$^2$, $\mu_\text{up}$ separates the multi phase area from Wigner phase area, when magnetic field is strong, $\mu_\text{up}$ separates multi phase area from another ordinary Nambu phase area. Of cause this classification is not as subtle as the phase diagram in \cite{allen}, but all these dividing lines are mathematically definable.

When magnetic field is below about $0.14$GeV$^2$, the upper boundary $\mu_\text{up}$ of chemical potential has an interesting irregular oscillation, this could be caused by quantum fluctuation around the critical point. However, a strong enough magnetic field would smear the fluctuation, that's why when $eB>0.14$GeV$^2$, $\mu_\text{up}$ has a regular smooth descending. Mathematically speaking, when $eB$ is small, there are several Landau levels play roles in the gap equation due to Heaviside step functions in $h(\sigma,\mu,eB)$, these levels cause the irregular solutions in Eq. (\ref{muup}), but when $eB$ is strong enough, that leaves us only the lowest Landau level ($\thetaf(\mu-\sigma)$ term in $h$), and it is much more regular and predictable. We can roughly estimate from which point on $\mu_\text{up}$ becomes regular, firstly, the phase transition line between chiral restored phase and chiral broken phase is around $0.3$GeV, so we assume $\mu=0.3$GeV, and now if we want only lowest Landau level involving in $h$, it requires $(\mu-\sqrt{2|q_\text{d}|eB})\le0$, which leads to $eB\ge0.135$GeV$^2$, it is pretty close to $0.14$GeV$^2$.

In this article we have proved the existence of multi phase, shown in Figs. (\ref{sigmu1}) and (\ref{sigmu2}). The case in Fig. (\ref{sigmu1}) belongs to MPA I of Fig. (\ref{phase}), in this area, not any chemical potential guarantees two valid solutions to the gap equation, only when chemical potential belongs to some specific intervals, the two valid solutions can be found. When we use numerical method to solve Eq. (\ref{gap3}), if one solution is close to the joint point of two different stairs in $h(\sigma,\mu,eB)$ such as $\sqrt{\mu^2-2n|q_\text{f}|eB}$, then maybe we could find another solution on the other side of the joint point. It probably has no mathematical equations to identify whether or not there is multi solution around $\sigma=\sqrt{\mu^2-2n|q_\text{f}|eB}$, we suspect this phenomenon happens whenever dynamic mass jumps from one stair to another with chemical potential changing, and the smaller dynamic mass is, the shorter intervals chemical potential need going through. The case in Fig. (\ref{sigmu2}) belongs to MPA II, as we can see in Fig. (\ref{phase}), this area spread from weak magnetic field to strong magnetic field, which means multi phase formed by ODM and UDM always exists as long as magnetic field is nonzero, this is understandable, because no matter how strong or weak the magnetic field is, from Nambu phase to Wigner phase, all particles have to pass through lowest Landau level, and this level causes multi phase.

\begin{figure}
 \centering
 \includegraphics[width=2.5in]{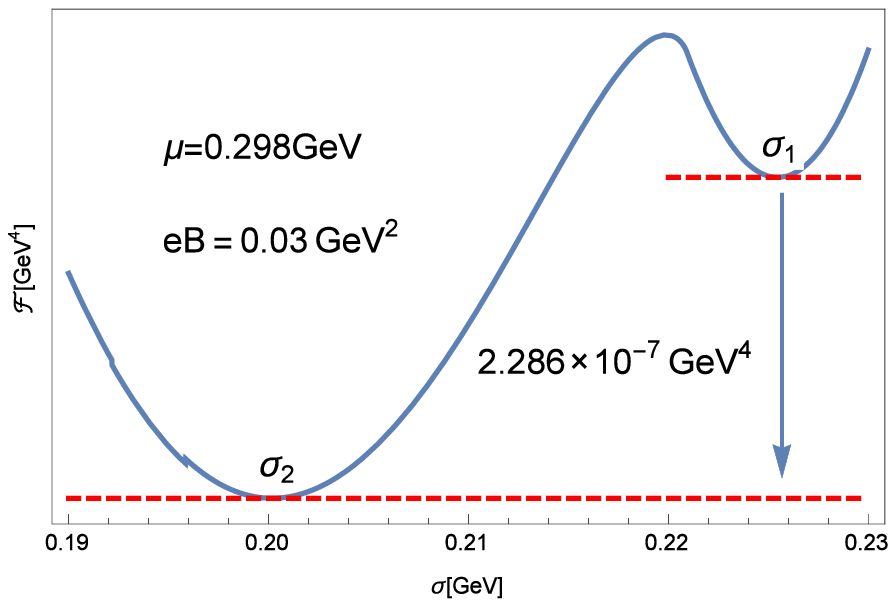}
 \includegraphics[width=2.5in]{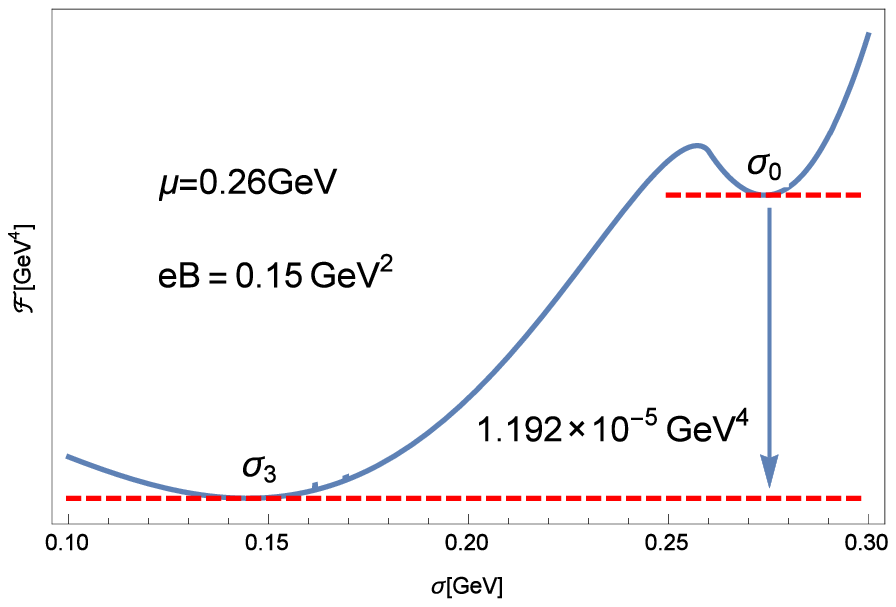}
 \caption{Examples of energy level transition. These two figures are corresponding to cases in Figs. (\ref{sigmu1}) and (\ref{sigmu2}) separately. The vertical axis represents free energy density $\mathcal{F}$, hence its nature units is GeV$^4$. $\sigma_1-\sigma_2\approx(0.022\text{GeV})^4$, $\sigma_0-\sigma_3\approx(0.059\text{GeV})^4$. Noticing only $\sigma_{0,1,2,3}$ have physical meaning, the other values of $\sigma$ in the intervals are virtual.\label{tran}}
\end{figure}

The physical effect of multi phase is energy level transition, we take the cases from Figs. (\ref{sigmu1}) and (\ref{sigmu2}) as examples, fix magnetic field and chemical potential at specific values which guarantee multi phase, treat $\sigma$ as free variable of free energy density $\mathcal{F}$ ($W=\int\mathcal{F}\,\mathrm{d}^4x$) from Eq. (\ref{free}), unsurprisingly there are two minimums of $\mathcal{F}$ in the intervals we choose, seen in Fig. (\ref{tran}). Because $\sigma_{1,2}$ (or $\sigma_{0,3}$ exist in the same external conditions (chemical potential and magnetic field), they can transfer to each other accompanied by energy absorption or radiation. If the absorption happens, absorbing photon for instance, some particles jump to higher Landau level (only one level higher for sure, because multi phase happens between adjacent levels), or vice versa. Interestingly when energy level transition happens, not only Landau level, but also dynamic mass changes, phenomenally speaking, the ``structure'' of quark matter has changed. We know if a thermal system is in equilibrium state with multiple phases, all phases must fulfill three kinds of equilibrium, thermal equilibrium (equal temperature), mechanical equilibrium (equal pressure) and diffusive equilibrium (equal chemical potential). In the multi phase case here, the quark matter remains at zero temperature (thermal equilibrium), with fixing chemical potential (diffusive equilibrium) and magnetic field (it is an external condition and space-time-independent), the free energy density (or pressure, $P=-\mathcal{F}$) is not equal, mechanical equilibrium is not fulfilled, therefore the particles in high energy state such as $\sigma_1$ in Fig. (\ref{tran}) will be pushed away by or transfer to particles in lower energy state $\sigma_2$, the multi phase state could not stably exist.

In this article, we employ zero temperature limitation and the ansatz $\Sigma=\sigma$, but as we mentioned at the beginning, a complete self energy should involve four kinds of condensate, if considering all these condensates, undoubtedly properties of phases for gap equations should be more complicate and abundant, in following work, we would like to do more detailed study, but as to a schematic view of quark matter at zero temperature, the ansatz in this article is adequate. The advantage of zero temperature limitation is that gap equation can be more ``clear'', and we are able to rely on mathematical tools to analyse gap equation, but when temperature goes to nonzero, we have to depend on numerical methods, it could be hard to find some properties of the solutions such as multi phase. We believe when temperature is low, multi phase phenomenon could still exist, because from zero temperature to low temperature, it is a continuous process, the properties of phases should be continuously varying. Also in following studies, we would like explore the cases at high temperature, although the particles' thermal motions at high temperature will smear many effects that can be found at low temperature, but if the multi phase is still there, that could be significant.

\appendix
\section{$\varepsilon$ Term in Propagator\label{prop}}
At zero temperature, there is always a $\varepsilon$ term in the  denominator of a particle propagator of momentum space, such propagator is generally have the form as $\frac{1}{p^2-m^2+i\varepsilon}$. Caused by relativistic causality, in momentum space, the $4$ dimensional integral of propagator should choose an appropriate contour in complex space, hence there comes the $\varepsilon$ term. But when chemical potential comes in, such term becomes chemical-potential-relevant, say, $i\varepsilon p_0(p_0-\mu)$, and it will essentially render a different contour in complex momentum space. In the above two examples, identifying $\varepsilon$ term is quite easy, just the regular canonical quantization routine to free particles. But what if the interaction terms come in or other external fields come in, how to identify the $\varepsilon$ term is a problem.

In this appendix, we have developed a convenient method to identify $\varepsilon$ term, which is inspired by \cite{dai}. The original idea is changing the `Hamiltonian' a bit in partition function
\begin{equation}
 \mathcal{Z}=\langle0|Te^{-i\int\hat H\,\mathrm{d}t}|0\rangle,
\end{equation}

In order to ensure the partition function is finite at infinite time, one can replace the Hamiltonian operator $\hat H$ with another complex version $\hat H'$,
\begin{equation}
 \hat H'=(1-i\eta)\hat H,\qquad\eta\to0^+,
\end{equation}
or replace Hamiltonian density instead
\begin{equation}
 \mathcal{H}'=(1-i\eta)\mathcal{H},\qquad H=\int\mathcal{H}\,\mathrm{d}^3x.
\end{equation}
Coincidently, when transferring this new Hamiltonian into Lagrangian, we can instantly get the correct particle propagator in momentum space, take free fermion for example
\begin{equation}
 \mathcal{H}'[\psi]=-(1-i\eta)\bar\psi(\gamma^ii\partial_i-m)\psi,
\end{equation}
\begin{equation}
 \mathcal{L}'=\frac{\partial\mathcal{L}}{\partial\dot{\psi}}-\mathcal{H}'=\bar\psi\hat D'\psi,\qquad\hat D=\gamma^0\partial_0+(1-i\eta)(\gamma^ii\partial_i-m).
\end{equation}

In momentum space, the inverse of operator $\hat D'$ is exactly the propagator we need
\begin{equation}
 \hat D'^{-1}=\frac{1}{\gamma^0p_0+(1-i\eta)(\gamma^ip_i-m)}=\frac{\gamma^0p_0+(1-i\eta)(\gamma^ip_i+m)}{p_0^2-(|\vec p|^2+m^2)(1-i\eta)^2}
 =\frac{\slashed{p}+m}{p^2-m^2+i\varepsilon},\quad\varepsilon=2(|\vec p|^2+m^2)\eta\to0^+.
\end{equation}
One can also prove this is effective to free boson propagator.

To demonstrate this method we have developed is valid, here we consider another example, the zero temperature and finite chemical potential case of free fermion propagator.
\begin{equation}
 \mathcal{H}'=-(1-i\eta)\bar\psi(\gamma^ii\partial_i+\mu\gamma^0-m)\psi,\qquad\mathcal{L}'=\bar\psi\hat D'\psi,
\end{equation}
\begin{equation}
 \hat D'=\slashed{\hat p}'-m',\quad \hat p_0'=\hat p_0+\mu',\quad\mu'=(1-i\eta)\mu,\quad \hat p_i'=(1-i\eta)\hat p_i,\quad m'=(1-i\eta)m.
\end{equation}
In momentum space, we have
\begin{equation}
 D'^{-1}=\frac{\gamma^0(p_0+\mu)+\gamma^ip_i+m}{(p_0+\mu)^2-\omega^2+i\varepsilon[\omega^2-\mu(p_0+\mu)]},\qquad\omega=\sqrt{|\vec p|^2+m^2},\label{pro0}
\end{equation}
this looks a bit different from what we need. First of all, in the momentum integral, we can always make a shift to $p_0$, $p_0+\mu\to p_0$. Secondly, in the denominator of Eq. (\ref{pro0}), $p_0^2-\omega^2$ implies the poles which require $p_0=\pm\omega$, we could use this relation to replace $\omega$ in $\varepsilon$ term, which gives us
\begin{equation}
 D'^{-1}\to\frac{\slashed{p}+m}{p_0^2-\omega^2+i\varepsilon p_0(p_0-\mu)}.
\end{equation}
That is what we need. Of cause in $\varepsilon$ term, the replacement of $\omega$ with $p_0$ seems a little undemanding. One can also prove rigorously from complex analysis that such replacement is legitimate.

\section{Eigenstate Method for Nonzero External Magnetic Field in Fermion Propagator\label{eigen}}
Assuming the fermion propagator with nonzero external magnetic field is
\begin{equation}
 \hat S_\text{f}=\frac{1}{/\kern-0.55em\hat\Pi^\text{f}-m},\qquad\hat\Pi^\text{f}_\mu=\hat p_\mu+q_\text{f}eA_\mu,\qquad A_\mu=(0,\frac{B}{2}x^2,-\frac{B}{2}x^1,0).
\end{equation}
Generally we need to deal with the cases such as $\Tr(\hat S_\text{f}\Gamma^a)$, where $\Gamma^a\in\{\gamma^\mu,\gamma^5,\gamma^5\gamma^\mu,\sigma^{\mu\nu}\}$. With the presence of external magnetic field, neither $|p\rangle$ nor $|x\rangle$ is $\hat S_\text{f}$'s eigenstate, therefore we need to find an appropriate representation through which we could avoid to confront the annoying noncommutative relation in $\hat S_\text{f}$ (eg. $[\hat\Pi_1,\hat\Pi_2]=-iq_\text{f}eB$). First, we try to scalarize the denominator of $\hat S_\text{f}$,
\begin{equation}
 \hat S_\text{f}=\frac{/\kern-0.55em\hat\Pi^\text{f}+m}{\hat p_0^2-\hat\Pi_\perp^2-\hat p_3^2-q_\text{f}eB\sigma^{12}-m^2},\qquad\hat\Pi^2_\perp=(\hat\Pi^\text{f}_1)^2+(\hat\Pi^\text{f}_2)^2.
\end{equation}
From the new propagator, we extract a series operators $(\hat p_0,\hat p_3,\hat\Pi_\perp^2)$ that commute with each others. Now we are able to define a eigenstate $|p_0,p_3\rangle\otimes|n,\lambda\rangle$ for these operators. $|p_0,p_3\rangle$ is obviously the eigenstate of $\hat p_{0,3}$,
\begin{equation}
 \hat p_{0,3}|p_0,p_3\rangle=p_{0,3}|p_0,p_3\rangle,
\end{equation}
$|n,\lambda\rangle$ is eigenstate of $\hat\Pi^2_\perp$,
\begin{equation}
 \hat\Pi^2_\perp|n,\lambda\rangle=(2n+1)|q_\text{f}|eB|n,\lambda\rangle,\qquad n\in\mathbb{N}^0,\qquad\lambda\in\mathbb{R},
\end{equation}
$\lambda$ is a free variable in the eigenstate, like $\theta$ in $e^{i\theta}$ as the free phase of wave function, it will not participate in the gap equation. Then for example the trace of $\hat S_\text{f}$ is
\begin{equation}
 \begin{split}
  \Tr\hat S_\text{f}&=\int\mathrm{d}p_0\,\mathrm{d}p_3\,\sum_{n=0}^{+\infty}\int_{-\infty}^{+\infty}\mathrm{d}\lambda\,\frac{\langle p_0,p_3;n,\lambda|/\kern-0.55em\hat\Pi^\text{f}+m|p_0,p_3;n,\lambda\rangle}{p_0^2-(2n+1)|q_\text{f}|eB-p_3^2-q_\text{f}eB\sigma^{12}-m^2}\\
  &=\frac{|q_\text{f}|eBm}{\pi}\int\frac{\mathrm{d}p_0\,\mathrm{d}p_3}{(2\pi)^2}\,\sum_n\frac{2-\delta_{0n}}{p_0^2-2n|q_\text{f}|eB-p_3^2-m^2}\int\mathrm{d}^4x
 \end{split}\label{gapb}
\end{equation}
for detailed deduction of above equations one can refer to the appendix in our previous work \cite{shi}.

\end{document}